\documentclass[11pt]{article}

\usepackage[svgnames]{xcolor}

\usepackage{scrextend}

\usepackage{natbib}

\usepackage{graphicx}
\usepackage{todonotes}
\usepackage{amssymb}  
\usepackage{amsmath}  
\usepackage{bm}

\usepackage[pdftex, colorlinks=true, linkcolor=blue, citecolor=blue]{hyperref}
\usepackage{helvet}

\usepackage{enumitem}
\usepackage{wrapfig}

\DeclareMathOperator*{\argmax}{arg\,max}

\begin{document}
\thispagestyle{empty}
\null\vspace{0.2\textheight}
\begin{center}{
\textbf{\Large Kernel Model Validation: How To Do It, And Why You Should Care}}\\
\vspace{2pt}
\large
Carlo Graziani and Marieme Ngom, Argonne National Laboratory, Lemont, IL\\

\end{center}
\vspace{5pt}
\textbf{\large Abstract}\hfill\\
Gaussian Process (GP) models are popular tools in uncertainty quantification (UQ) because they purport to furnish functional uncertainty estimates that can be used to represent model uncertainty. It is often difficult to state with precision what probabilistic interpretation attaches to such an uncertainty, and in what way is it calibrated.  Without such a calibration statement, the value of such uncertainty estimates is quite limited and qualitative.  We motivate the importance of proper probabilistic calibration of GP predictions by describing how GP predictive calibration failures can cause degraded convergence properties in a target optimization algorithm called Targeted Adaptive Design (TAD). We discuss the interpretation of GP-generated uncertainty intervals in UQ, and how one may learn to trust them, through a formal procedure for covariance kernel validation that exploits the multivariate normal nature of GP predictions. We give simple examples of GP regression misspecified 1-dimensional models, and discuss the situation with respect to higher-dimensional models.\\
\vspace{2pt}

\section{Introduction}
Gaussian Process (GP) models are a family of non-parametric statistical methods within the broader family of kernel methods.  They are very popular in the machine learning (ML) research community, and have proved useful in image reconstruction \citep{he2011single}, X-ray tomography \citep{li2013bayesian,dasgupta2023simultaneous}, surrogate modeling \citep{gramacy2020surrogates}, emulation of expensive-to-compute simulation output \citep{sacks1989designs,bastos2009diagnostics}, black-box function optimization \citep{schonlau1998global,graziani2024targeted}, optimal experimental design \citep{zhang2016adaptive}, and many other applications.

Formally, a Gaussian Process is defined as a collection of random variables, any
finite number of which have a joint Gaussian distribution \citep[$\S2.2$]{williams2006gaussian}. Informally, it is useful to think of a GP as a generalization of the finite-dimensional multivariate normal (MVN) distribution to infinite-dimensional (Hilbert) spaces of functions. For example, if one thinks of the Fourier series representation of a function, wherein the sin and cosine functions are basis vectors and the Fourier coefficients are the components of the function along those vectors, then one is not seriously misled by thinking of an MVN over the coefficients as an example of a GP. Just as an MVN is characterized by its mean vector $\bm{\mu}$ and symmetric-positive-definite (SPD) covariance matrix $\bm{C}$, a GP over some space $\Omega\subset\mathbb{R}^N$ is characterised by a \emph{mean function} $\mu(\bm{x})$ and an SPD \emph{covariance kernel} $K(\bm{x},\bm{x}^\prime)$, $\bm{x},\bm{x}^\prime\in\Omega$. Intuitively, one may think of the vector/matrix indices being superseded by continuous indices $\bm{x},\bm{x}^\prime$.

GPs model functions as random variables, sampled from a space of functions. The characteristics of the function space are in fact dictated by the choice of covariance kernel $K(\cdot,\cdot)$: by choosing different functional forms for $K()$, one may choose to sample functions of chosen continuity, order of differentiability (i.e. degree of smoothness), scale, periodicity, etc. Despite the restrictive requirement that $K()$ should be SPD as an integral kernel, there are many possible choices of kernel forms, and new kernels may be composed from old ones in various ways, to suit the nature of the data under analysis  \citep[Chapter 4]{williams2006gaussian}.

Suppose that we have an unknown function $f(\bm{x})$ from which we have obtained a number of samples $f(\bm{x}_l)$, $l=1,\ldots,N$.  It is possible to infer the properties of $f(\bm{x})$ by modeling it as governed by a GP with some mean function $\mu(\bm{x})$ and covariance kernel $K(\bm{x},\bm{x}^\prime)$---we write this $f()\sim\mathcal{GP}(\mu(),K())$. If the functional forms of $\mu()$ and $K()$ are parametrized, it is possible to estimate those parameters from the data by maximizing the likelihood, a process known as \emph{training}. Further, it is possible to \emph{predict function values at other locations, with uncertainty estimates on the prediction}. The prediction is, in effect, the conditional probability of the function value at the new location, conditioned on the observed function values.  If the observations $f(\bm{x}_l)$ were contaminated by Gaussian additive noise with variance $\sigma^2_l$, the effect of this noise can also be incorporated into the prediction  
\citep[Chapter 2]{williams2006gaussian}.

To make this concrete: Suppose that we have a set of points $\bm{x}^*_k$, $k=1,\ldots,M$ at which we would like predictions.  Let us denote the ensemble of $\bm{x}_l$ by $\bm{X}$, the ensemble of $\bm{x}^*
_k$ by $\bm{X}^*$, the observed function values by a vector $\bm{f}$, and the unknown function values at the points $\bm{X}^*$ by a vector $\bm{f}^*$.  We compute matrices $\bm{K}\equiv K(\bm{X},\bm{X})\in\mathbb{R}^{N\times N}$, $\bm{K}^*\equiv K(\bm{X}^*,\bm{X})\in\mathbb{R}^{M\times N}$, and $\bm{K}^{**}\equiv K(\bm{X}^*,\bm{X}^*)\in\mathbb{R}^{M\times M}$, $\bm{\mu}\equiv\mu(\bm{X})\in\mathbb{R}^N$, $\bm{\mu}^*\equiv\mu(\bm{X}^*)\in\mathbb{R}^M$, by which we mean that these arrays are computed by evaluating the kernel and mean functions at all the corresponding points in $\bm{X},\bm{X}^*$.  We also represent the measurement noise by a diagonal covariance matrix $\bm{N}\equiv\textrm{Diag}(\sigma_1^2,\ldots,\sigma_N^2)\in\mathbb{R}^{N\times N}$. Then the prediction takes the form
\begin{eqnarray}
\bm{f}^*|\bm{f},\bm{X},\bm{X}^* &\sim& \mathcal{N}(\bm{\mu}_{pred},\bm{K}_{pred}),\nonumber\\
\bm{\mu}_{pred} &\equiv& \bm{\mu}^* + 
\bm{K}^*\left(\bm{K}+\bm{N}\right)^{-1}\left(\bm{f}-\bm{\mu}\right)
\label{eq:predmean}\\
\bm{K}_{pred} &\equiv& \bm{K}^{**}-\bm{K}^*\left(\bm{K}+\bm{N}\right)^{-1}{\bm{K}^*}^\top.\label{eq:predcov}
\end{eqnarray}

This is to say that the prediction comes in the form of an MVN, with computable mean vector (a linear function of the observed function values) and a computable covariance (which depends on the observation locations, but not directly on the observations).  The covariance can be seen to be reduced in its eigenvalues from the prior covariance $\bm{K}^{**}$ by a positive-definite term, which corresponds to reduced uncertainty due to the conditioning observations.

\begin{figure}[t]
\begin{center}
\includegraphics[width=0.85\textwidth]{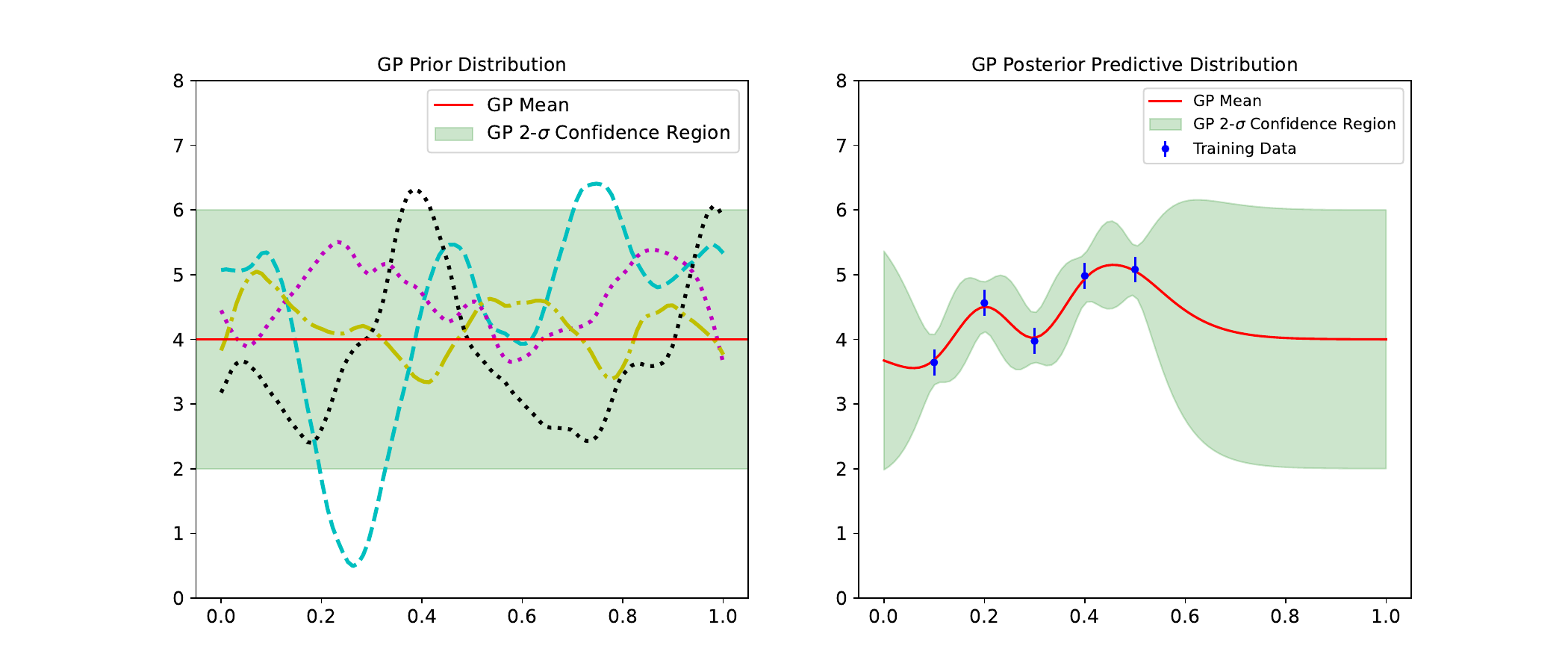}
\end{center}
\caption{Illustration of GP prediction. Left panel: Prior GP distribution. The red horizontal line is the (constant) mean function, while the green band represents the diagonal elements of the covariance (i.e. the variance). The curves shown are random samples from this GP distribution. Right panel: Conditional GP distribution. The red points are noisy measurements of the function, and the red line and green band show the predictive mean and variance.\label{Fig:gp_ill}
\vspace{5pt}}
\end{figure}

Figure \ref{Fig:gp_ill} illustrates this prediction process. The left panel shows a prior GP distribution over functions in $\mathbb{R}$, including some functions sampled from the distribution.  The right panel shows the effect of conditioning on some noisy data: the predictive mean now adjusts to the data, providing an interpolating curve, while the covariance diagonal elements (the variances) shrink with respect to their prior values, particularly in the vicinity of the data.  As the prediction points leave the support of the data, the mean and variance relax back to a kernel-weighted mean and a kernel-weighted variance of the data points, respectively. In terms of our intuitive picture of an MVN in a space of functions, this conditioning is analogous to conditioning a $D$-dimensional MVN on observed values of $m$ of its coordinates: the result is a new MVN over the remaining $D-m$ coordinates, with a shifted mean and reduced covariance, according to formulae very similar to those in Equations (\ref{eq:predmean},\ref{eq:predcov}).

This example illustrates the reason that GP modeling is so appealing to practitioners of Uncertainty Quantification (UQ): GPs allow one to combine uncertainties due to observational noise and due to an uncertain model into a single uncertainty on predictions. This is in contrast to non-parametric methods such as optimal recovery \citep{wendland2004scattered} or splines \citep{greville1969introduction}, which produce interpolating data smoothings, but no estimates of model uncertainty.

There is a fly in the ointment, however: different kernel choices can result in quite different predictive covariances.  How to know which is correct, and what ``correctness'' even means when applied to such a covariance?  What sort of probability coverage should one attach to a Bayesian credible region obtained by GP prediction, and how should that coverage be validated?  One frequently finds in the literature examples of GP models that are ``validated'' because the mean-squared error (MSE) of the mean prediction from some held-out validation data is ``small,'' but this is not satisfactory as a model validation statement, because it appeals to no principled notion of what level of smallness is acceptable. In any event, such model validations say nothing about the interpretation of predictive uncertainty, and leave some mystery concerning how one is supposed to use or interpret such uncertainty.

In this work we discuss an approach to assessing kernel choice that is based on the quality of probabilistic predictions on held-out test data, or on data acquired after GP regression has been performed.  We discuss the use of the \emph{Mahalanobis distance} as a quantitative indicator of fit quality, referring to our use of this indicator in our previous work \citep{graziani2024targeted} to motivate its introduction.  We introduce new methodology based on diagonalization of the predictive covariance to produce I.I.D. normal variates, whose CDFs may be compared directly to the uniform distribution that they should be sampled from if the kernel model is adequate. We illustrate the updated methodology with numerical experiments using misspecified kernel models, and use them to give intuitively reasonable analyses of model misspecification.

The subject of kernel misspecification in GP modeling has received a good deal of attention in the literature, although in regression settings it is usually the interpolating mean function alone whose properties have been studied.  \cite{wynne2021convergence} studied the convergence rates of GP regression means under likelihood and smoothness misspecification.  \cite{wang2022gaussian} studied the effect of kernel smoothness misspecification on GP and kernel ridge regression, again with a focus on convergence rates of mean functions. \cite{bogunovic2021misspecified} discussed probabilistic regret bounds with misspecified kernels in kernelized bandit problems. \cite{stephenson2021measuring} examined decision robustness to kernel choice, in terms of constraints on families of stationary kernels implemented by bounding the departure from a reference kernel in spectral space. Our contribution departs from these efforts in that we focus on regression, but broaden the discussion of misspecification from convergence of mean functions to the quality of probabilistic predictions. This is the aspect that seems to us most relevant to UQ applications of GP modeling, and specifically to the question of how one is to use and interpret the predictive probability confidence bounds issued by such models.

\section{A Cautionary Tale: Targeted Adaptive Design}\label{sec:TAD}

The importance of some form of data-driven kernel model validation was driven home for the authors during the course of the development of the \emph{Targeted Adaptive Design} (TAD) algorithm \citep{graziani2024targeted}.  TAD is a black-box function target optimization method intended to assist advanced manufacturing and materials design.  In such design problems, one has a set of possible experimental settings, some product or material features that one wishes to produce within some tolerances, and an unknown, computationally- or experimentally-expensive to determine response mapping from settings to features.  TAD attempts to locate settings such that the resulting estimated features have uncertainties that fit within the tolerance box. More formally: given a design space $\Omega\subset\mathbb{R}^D$, a feature space $\mathbb{R}^F$, an unknown function $f:\Omega\rightarrow\mathbb{R}^F$, some set of target features $f^{(T)}_k$, $k=1,\ldots,F$ and a set of feature tolerances $t_k$ (so that the ``tolerance box'' is $f^{(T)}_k\pm t_k$, $k=1,\ldots,F$), TAD creates a GP surrogate function $\hat{f}()$ intended to approximate $f()$,  and trains it on existing data points, $f(\bm{X}^{(1)})$.  This model predicts the value of $\hat{f}(\bm{x}_2)$ at new points $\bm{x}^{(2)}$, including uncertainty, according to the predictive distribution formulae given above. TAD attempts to locate a point $\bm{x}^{(T)}$ such that all the 1-D GP uncertainties in $\hat{f}(\bm{x}^{(T)})$ fit in the tolerance box about $f^{(T)}$.

\begin{figure}[t]
\begin{center}
\includegraphics[width=0.5\textwidth]{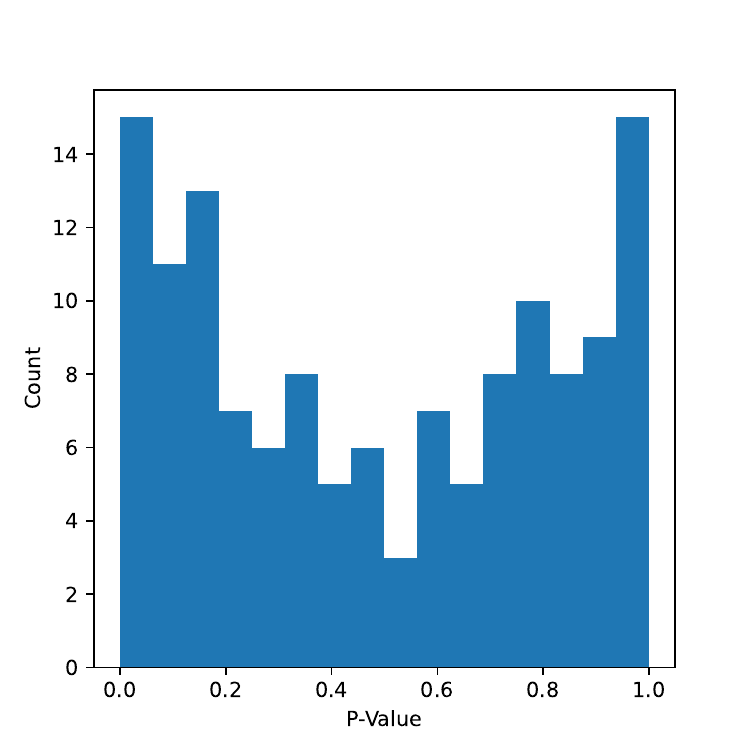}
\end{center}
\caption{$P$-values obtained from a TAD run with a fixed, simplistic GP model.\label{Figure:TAD-PVals}}
\vspace{5pt}
\end{figure}

The algorithm does this by proposing a set of $N_2$ probe points $\bm{X}^{(2)}$ together with a proposed target point $\bm{x}^{(T)}$, and at each stage solving the optimization problem
\begin{equation}
\argmax_{\bm{X}^{(2)},\bm{x}^{(T)}} E_{\hat{f}(\bm{X}^{(2)})|\bm{X}^{(2)},\bm{X}^{(1)},f(\bm{X}^{(1)})}\left\{
\log\pi_{\mathcal{N}(\bm{\mu}_{pred},\bm{K}_{pred})}
\left(\hat{f}(\bm{x}^{(T)})=f^{(T)}\right)
\right\},\label{eq:TAD_AF}
\end{equation}
where $\pi_{\mathcal{N}(\bm{\mu}_{pred},\bm{K}_{pred})}()$ is the MVN density of the predictive distribution  $f(\bm{x}^{(T)})|f(\bm{X}^{(1)}),f(\bm{X}^{(2)})$ with mean $\bm{\mu}_{pred}$ and covariance $\bm{K}_{pred}$. The expectation over $f(\bm{X}^{(2)})$ is necessary, because the function values at the probe points are only acquired after the solution of the optimization problem.  The objective in Equation (\ref{eq:TAD_AF}) is attracted to regions in which the surrogate $\hat{f}(\bm{x}^{(T)})$ is known, with increasing certainty, to resemble $f^{(T)}$, and repelled from regions where $\hat{f}(\bm{x}^{(T)})=f^{(T)}$ is unlikely.  At the end of each optimization, the data $f(\bm{X}^{(2)})$ is acquired, $\bm{X}^{(2)}, f(\bm{X}^{(2)})$ are added to the in-the-can data $\bm{X}^{(1)}, f(\bm{X}^{(1)})$, and the process begins again.  The algorithm terminates if it finds a solution or if it concludes, based on an information criterion, that a solution is unlikely to exist.

Early versions of TAD suffered from an odd pathology: in testing on artificial functions $f()$ with a known solution $\bm{x}^{(T, true)}$, the proposed target point $\bm{x}^{(T)}$ would at first approach $\bm{x}^{(T, true)}$, then circle it for many iterations converging very slowly if at all.  Further investigation showed that the acquired function values $f(\bm{X}^{(2)})$ were very unlikely in the distribution $\mathcal{N}(\bm{\mu}_{pred},\bm{K}_{pred})$.  The acquisition function in Equation (\ref{eq:TAD_AF}) only behaves as expected if the normal model for $f(\bm{X}^{(2)})$ is reasonably accurate. Since it was not a reasonable approximation it was not possible to achieve convergence.

The diagnosis of poor fit was accomplished by computing the \emph{Mahalanobis distance}:
\begin{equation}
\chi^2_M\equiv \left[\bm{f}^{(2)}-\bm{\mu}_{pred}\right]^\top{\bm{K}_{pred}}^{-1}
\left[\bm{f}^{(2)}-\bm{\mu}_{pred}\right].\label{eq:Mahalanobis}
\end{equation}
If the data is in fact distributed as $\bm{f}^{(2)}\sim\mathcal{N}(\bm{\mu}_{pred},\bm{K}_{pred})$, then $\chi^2_M$ is distributed as a $\chi^2$ distribution with $p$ degrees of freedom, where $p$ is the dimensionality of $\bm{f}^{(2)}$ \citep{johnson2002applied,ghorbani2019mahalanobis}. The $\chi^2_M$  value obtained for the $\bm{f}^{(2)}$ samples acquired on the failed approaches to the target point were not plausible in this distribution. Figure \ref{Figure:TAD-PVals} illustrates the problem: the figure shows a histogram of the $P$-values computed from the $\chi^2_M$ during a run that required 136 iterations to achieve convergence. This distribution, which should be uniform if the model describes the data adequately, exhibits very clear non-uniformity, with peaks at small and large $P$-values.  The predictive model was inaccurate, in that the principal axes of its confidence ellipsoids did not conform to the data well, resulting in slow convergence.

The solution to this problem turned out to be automatic kernel model reform: the algorithm would compute $\chi^2_M$ after acquiring new data.  Repeated anomalous $P$-values would trigger a change in the kernel.  The kernel model was an additive superposition of terms with different parameters.  The ``reformed'' kernel model would receive an additional such term, randomly initialized, and the parameter training would be repeated on this more flexible model.  Then the optimization would be restarted. This change secured the algorithm's ability to converge to the target solution in 18 iterations.

The lesson that we drew from this experience is that poor kernel model choice can have serious consequences---in our case, convergence failures---when the probabilistic nature of a GP's predictions are taken seriously.  Where only the predictive mean smoother/interpolator is of interest, one can perhaps get away with a na\"ive kernel choice. But if one is engaged in the enterprise of properly quantifying uncertainty from GP-based inference, such a choice is obviously unacceptable.

\section{Tests With Misspecified Models}

Let us make a more systematic clean-room study of the effects of kernel model misspecifcation, so as to arrive at good methods to detect and control such misspecification.

\begin{figure}[t]
\begin{center}
\includegraphics[width=0.42\textwidth]{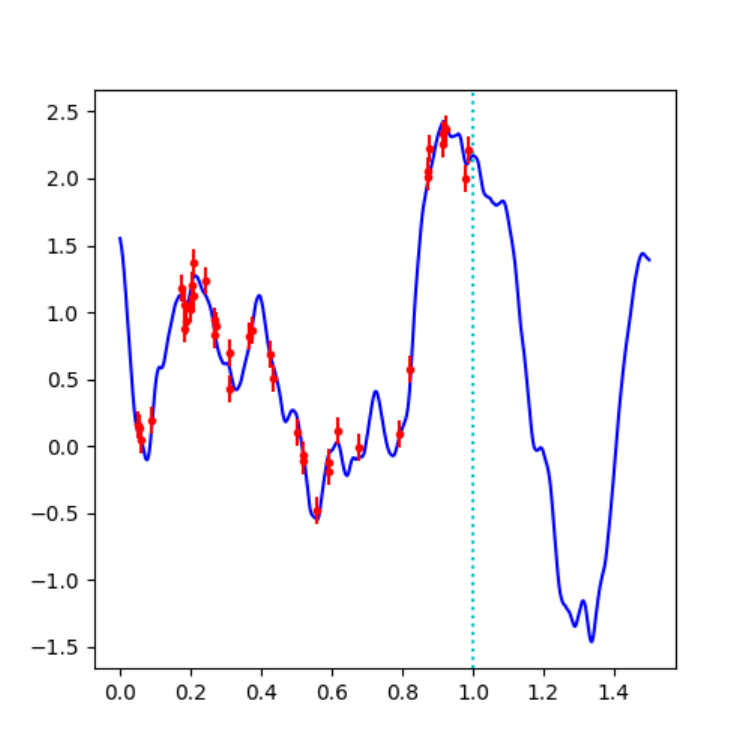}
\includegraphics[width=0.42\textwidth]{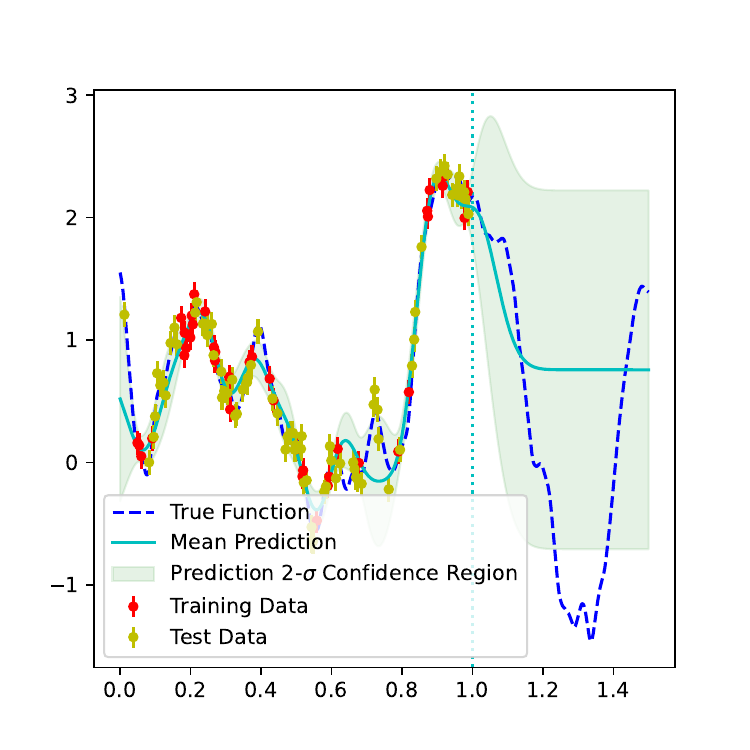}
\caption{\label{Fig:badmodel}Left panel: A function sampled from a GP with mean zero and a $\nu=1.5$ Matern covariance kernel (blue line) and some points sampled noisily from the curve. The model yields rough (at most once-differentiable) functions. Right panel: a GP model with a squared-exponential covariance kernel (which models very smooth $C^\infty$ functions) is fit to the red training points. Its mean is shown by the cyan solid line, while the 2-$\sigma$ credible interval is shown by the cyan filled band.  The yellow points are 80 test points, held out from training. The true function is displayed by the blue dashed line.
\vspace{5pt}}
\end{center}
\end{figure}

\subsection{A Badly Misspecified Model}

Figure \ref{Fig:badmodel} shows just such a simple test. In the left panel we see a function (the blue line) sampled randomly from a zero-mean GP with a $\nu=1.5$ Matern covariance kernel \citep[Chapter 4]{williams2006gaussian}. This kernel governs a distribution over once-differentiable ($C^1$) functions, which have quite rough behavior, illustrated by the sampled curve. The red points are function values sampled from this curve, to which some Gaussian noise is added (indicated by the errorbars).

The right-hand  panel shows a GP model fit.  The covariance kernel is a squared-exponential (also sometimes called an ``Radial Basis Function'' or RBF kernel). This covariance governs a distribution of infinitely-smooth ($C^\infty$) characterized by a single scale length. In other words, this is a very misspecified model, not at all appropriate for the type of functional sample data in this example. The predictive mean is shown by the solid cyan curve, while the $2-\sigma$ credible interval is shown by the filled cyan band. The true function is shown as a dashed blue line.  In addition to the red training points, 80 new test points that were held out of training are shown in yellow. The predictive mean from the fit does an adequate job of staying close to the test points, although some large deviations are perceptible.  However, the predictive covariance does a terrible job of modeling the scatter in the test points: we observe $\chi^2_M=129$ for DOF=80, a $P$-value of $4.3\times 10^{-4}$. Clearly, the predictive uncertainties are quite wrong with this model, and would be worthless for UQ purposes.


The Mahalanobis distance appears to be a useful tool for diagnosing model misspecification. However, it is really a coarse summary, producing a single $P$-value from the $N=80$ held-out data points.  This could be a cause for concern:  $\chi^2_m$ is a quadratic function of all the mean-deviation residuals in the data. How do we know that a ``good'' $P$-value (e.g. $P=0.2$) doesn't derive from a mixture of too-large and too-small residuals, averaging out to a reasonable $\chi^2_M$?  Both excessively-large residuals (i.e. tiny $P$-values) and excessively-small residuals (i.e. $P$-values very close to 1) are indications of failure of the normal model to describe the data properly. It would be valuable to have a tool that is sensitive to such mixed model pathologies.

We can construct such a method by noting that $\bm{K}_{pred}$, being an SPD matrix, may be diagonalized by an orthogonal transformation, and the residuals may be projected along its normal modes:
\begin{equation}
\bm{O}^\top\bm{K}_{pred}\bm{O}=\textrm{Diag}(s_1^2,\ldots,s_N^2) \quad;\quad
\bm{d}\equiv \bm{O}^\top\left[\bm{f}-\bm{\mu}_{pred}\right].
\label{eq:norm_decomp}
\end{equation}
With this decomposition, it is clear from the diagonality of the transformed covariance that if the GP model describes the data adequately, then the components $d_k$ of $\bm{d}$ satisfy
\begin{equation}
e_k\equiv\frac{d_k}{s_k}\sim \mathcal{N}(0,1),
\label{eq:iid}
\end{equation}
that is, the $e_k=d_k/s_k$ are i.i.d. standard normal random variables.

We can exploit this fact to make a simple visualization of model adequacy. We compute the standard normal survival distribution function, which is related to the cumulative distribution function (CDF) values of the observed $e_k$ by 
\begin{equation}
p_k=1-\textrm{CDF}(e_k)=\frac{1}{2}\left[1-\textrm{erf}(e_k/\sqrt{2}) \right],
\end{equation}
where $\textrm{erf}(z)$ is the error function.
If the GP model describes the data adequately, the $p_k$ should be i.i.d. uniformly, $p_k\sim\mathcal{U}(0,1)$.

\begin{figure}
\begin{center}
\includegraphics[width=0.44\textwidth]{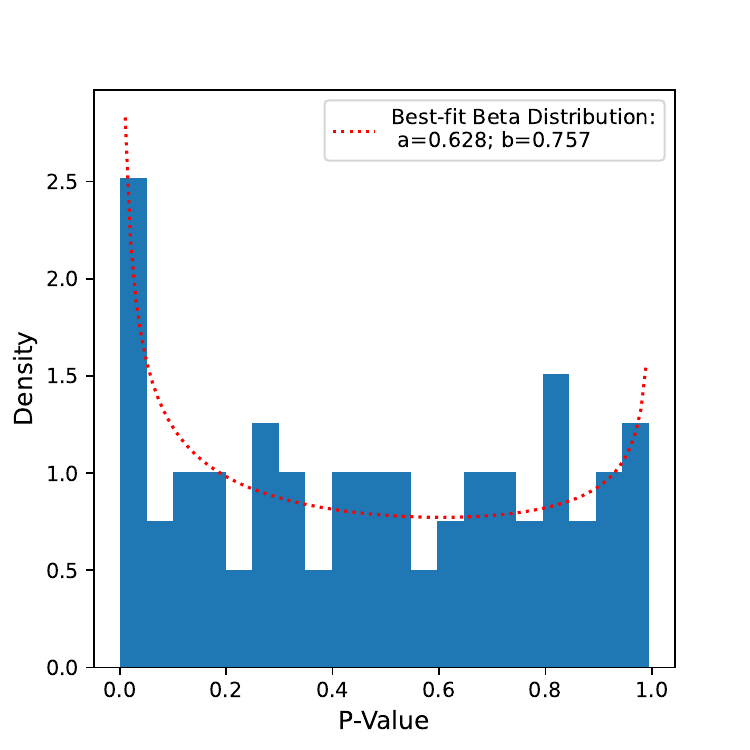}
\includegraphics[width=0.52\textwidth]{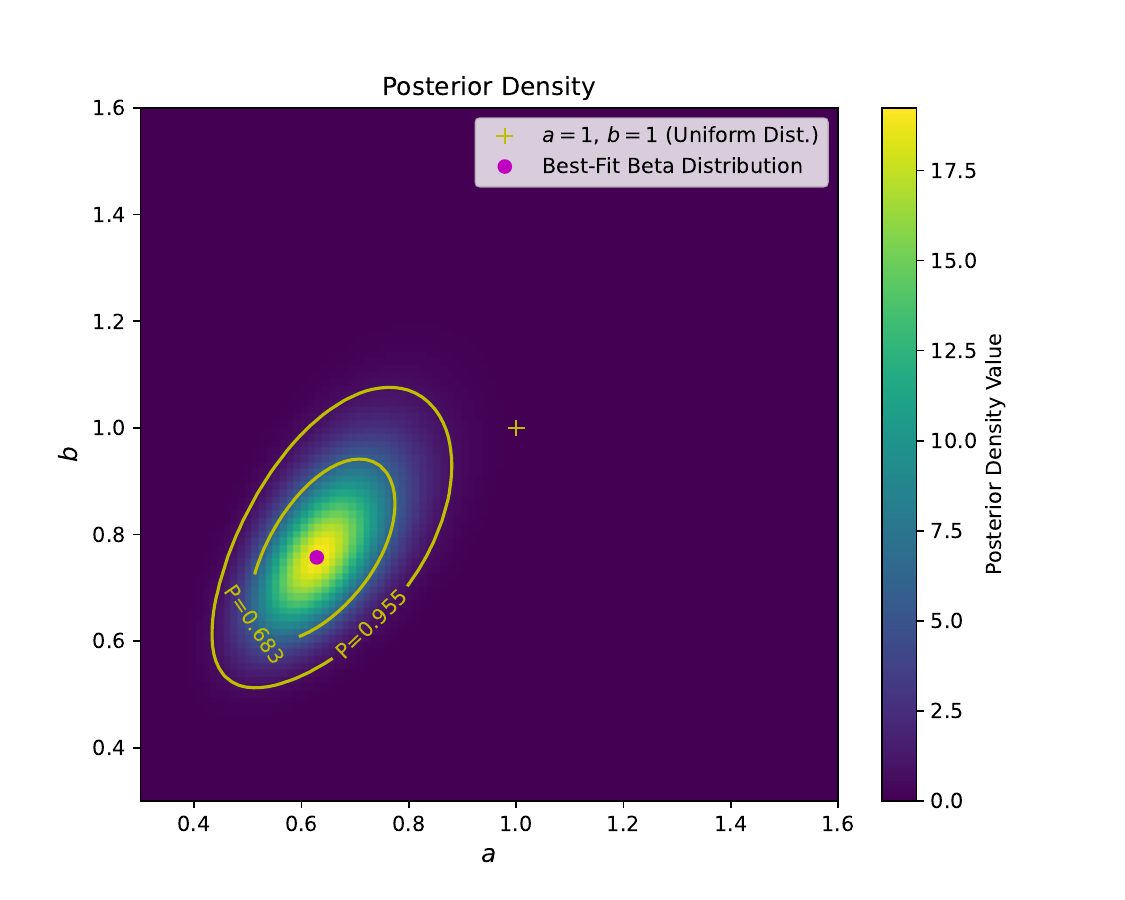}
\end{center}
\caption{\label{Fig:normdecomp-badmodel}Left panel: Binned density histogram of the residual survival probabilities $p_k$ from the GP fit of Figure \ref{Fig:badmodel}.  The dotted red line displays the best-fit Beta distribution to the (unbinned) data. Right panel: The Bayesian posterior over the two Beta distribution parameters, assuming a uniform prior.\vspace{5pt}}
\end{figure}

Figure \ref{Fig:normdecomp-badmodel} shows how the  $p_k$ may be put to good use. The left panel shows a binned density histogram of the residual survival probabilities $p_k$, which appears to place too much weight at low $p_k$ (and possibly at high $p_k$ as well). In order not to rely solely on visual impressions, we can fit a Beta distribution density to this data.  The Beta density is given by \citep[Chapter 8]{forbes2011statistical}
\[
\pi_{\beta(a,b)}(p)=\frac{\Gamma(a+b)}{\Gamma(a)\Gamma(b)}p^{a-1}(1-p)^{b-1},
\]
with $a>0$, $b>0$. In general, the Beta distribution is useful here because it can model many common densities on $(0,1)$, including the uniform distribution $\mathcal{U}(0,1)$, which results from choosing $a=1,b=1$.

Since the $p_k$ should be independent if the model is valid, the likelihood of the unbinned data assuming a Beta distribution is
\[
L(a,b)=\prod_{k=1}^N\pi_{\beta(a,b)}(p_k).
\]
A valid kernel model should result in a likelihood concentrated near the uniform distribution parameters $a=1, b=1$.

We may optimize this likelihood with respect to the parameters $a,b$. The best-fit model, shown as the dotted red line in the left panel of Figure \ref{Fig:normdecomp-badmodel}, corresponds to $a=0.628$, $b=0.757$. The fit certainly appears to favor a model with a very heavy concentration near $p_k=0$, and some concentration near $p_k=1$.

To make these observations more quantitative, we may trade in the likelihood for a Bayesian posterior density over the parameters $a,b$. We assume a uniform prior density over the parameters.  Since the parameter space is only 2-dimensional, we can perform all required integrals by simple quadratures of the binned posterior, rather than going to the trouble of performing an MCMC to obtain inferences. This posterior density is exhibited in the right panel of Figure \ref{Fig:normdecomp-badmodel}.  The colormap gives posterior density values. The uniform model $a=1$, $b=1$ is shown as the +-sign in the figure.  The best-fit parameters are shown by the red dot, and 68.3\% and 95.5\% iso-posterior credible regions are shown as closed yellow contours.  The uniform model lies well outside these contours: in fact, the iso-posterior credible region whose boundary crosses the uniform model contains a coverage probability $1-1.3\times 10^{-4}$. This analysis would lead us to conclusively exclude the possibility that the kernel model employed here furnishes an adequate model of this data.

\begin{figure}[t]
\begin{center}
\includegraphics[width=0.30\textwidth]{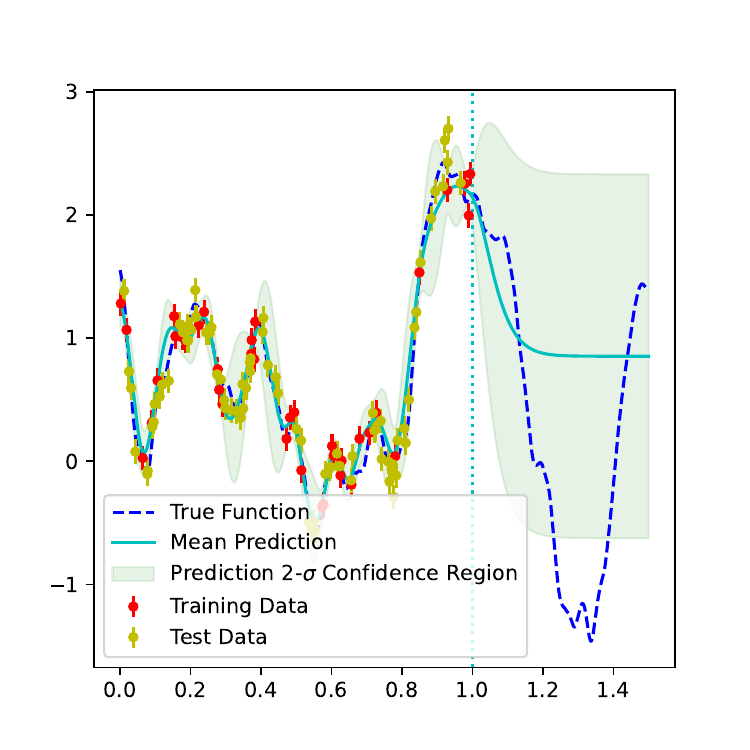}
\includegraphics[width=0.30\textwidth]{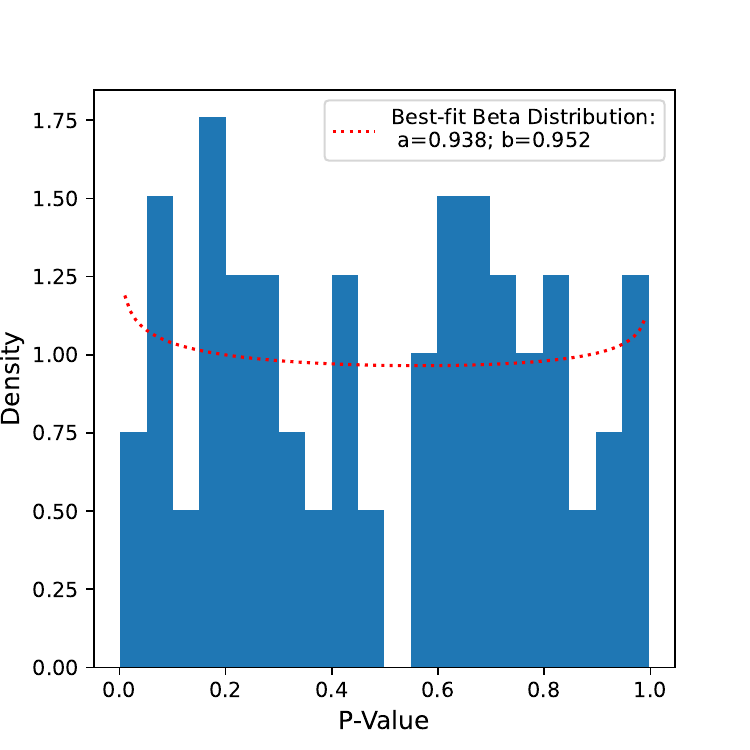}
\includegraphics[width=0.36\textwidth]{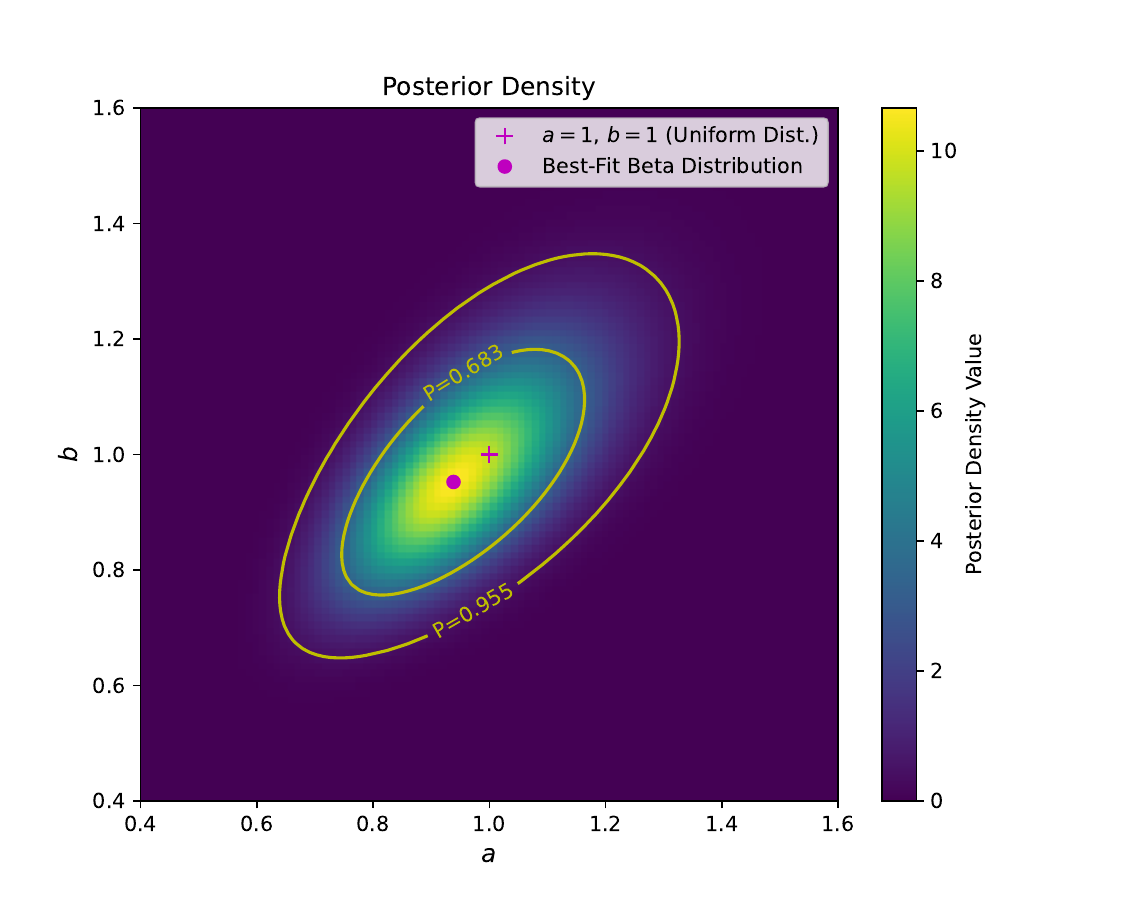}
\end{center}
\caption{\label{Fig:better-test} Left panel: A GP fit to the same data as in Figure \ref{Fig:badmodel}, with a $\nu=2.5$ Matern covariance. Middle panel: Binned density histogram of the $p_k$. Right panel: posterior density over the parameters of a Beta distribution fit to the $p_k$.\vspace{5pt}}
\end{figure}

\subsection{Less Misspecified Models}

Let us repeat the experiment with a less na\"ive model, one that we expect to be less misspecified for this data than a squared-exponential. We will instead use a $\nu=2.5$ Matern covariance.  This covariance kernel governs a distribution over at most twice-differentiable ($C^2$) functions. These are smoother than the $C^1$ function space from which the data was generated, so that the model is still somewhat misspecified. But we expect the degree of misspecification to be less than was the case for a squared-exponential kernel.

Figure \ref{Fig:better-test} shows the result of the repeated analysis with the improved kernel model. Comparing the left panel of the figure with the middle panel of Figure \ref{Fig:badmodel} we can already perceive some improvement in the ability of the mean function to track the held-out data.  A quantitative measure of the improvement is that the Mahalanobis distance resulting from the fit is $\chi^2_M=85.8$ for DOF=80, which gives a very respectable $P$-value of 0.307.

As discussed earlier, a good Mahalanobis $P$-value is not dispositive of an adequate model fit, since too-large and too-small model residuals can conspire to produce a deceptively reasonable value of $\chi^2_M$.  We repeat the Bayesian
analysis from the previous subsection to test whether this is the case here. The results are shown in the middle and right panels of Figure \ref{Fig:better-test}. We see in the middle panel that the histogram is no longer as strongly concentrated to low values of $p_k$, as confirmed by the best-fit values of $a=0.938$, $b=0.952$, which are now much closer to $a=1,b=1$.  In the right panel, we see that the uniform distribution $a=1,b=1$ is quite plausible in the Bayesian posterior. The iso-posterior contour that crosses this point has a coverage probability of 0.095, confirming the plausibility of this model.

\begin{figure}[t]
\begin{center}
\includegraphics[width=0.30\textwidth]{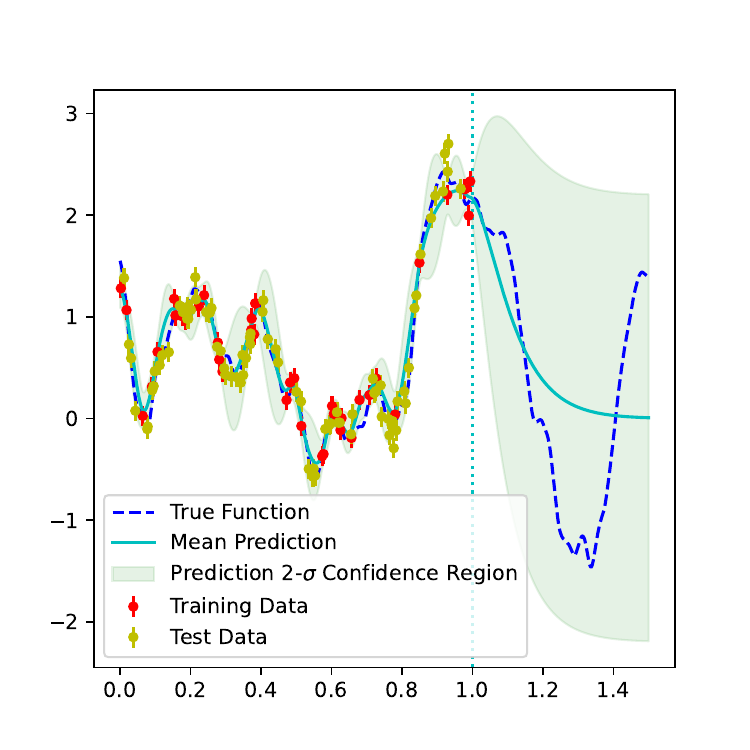}
\includegraphics[width=0.30\textwidth]{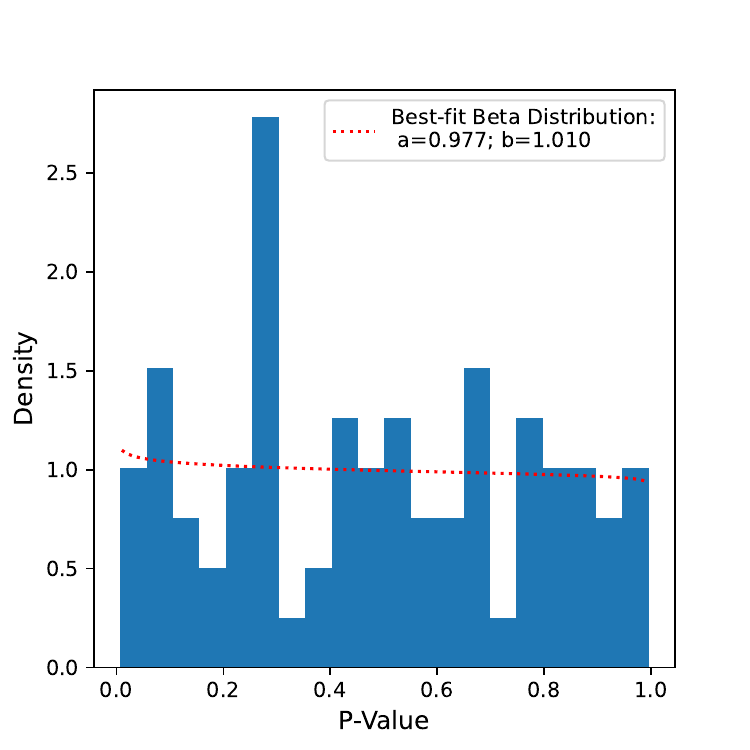}
\includegraphics[width=0.36\textwidth]{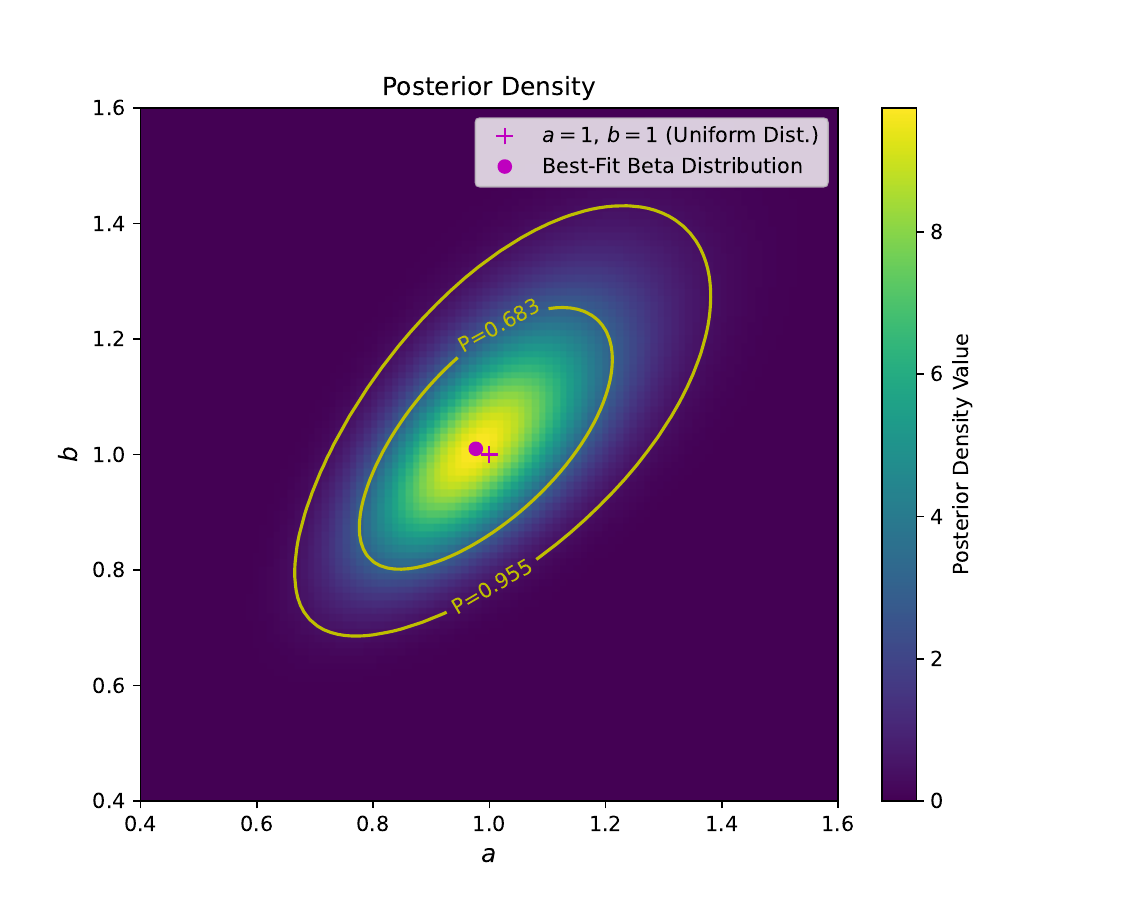}
\end{center}
\caption{\label{Fig:perfect-test} Similar to Figure \ref{Fig:better-test}, but for the ``perfect'' model with a $\nu=1.5$ Matern kernel.
}
\vspace{5pt}
\end{figure}


Finally, let's look at what happens when we use the $\nu=1.5$ Matern covariance, zero-mean model that was actually used to generate the data in the first place. The results are shown in Figure \ref{Fig:perfect-test}. The Mahalanobis distance resulting from the fit is $\chi^2_M=80.7$ for DOF=80, which gives a  $P$-value of 0.456.  The results of the Bayesian analysis are displayed in the middle and right panels of the Figure. The histogram is now very close to uniform, with best-fit Beta parameters $a=0.977$, $b=1.010$, and the iso-posterior contour that crosses the $a=1$, $b=1$ point has a coverage probability of 0.035. The model performance is thus excellent, as expected.
In the context of GP modeling, this is of course unrealistic model fidelity, since the ``correct'' GP model is unknowable for real-world data, and such data is rarely generated from a GP anyway, so that we can at best regard GP models as approximations to the data-generating distribution.  The lesson here is not that we should search for the true GP model, but rather that we should be careful to reject models that clearly have no predictive power with respect to the data.

\section{Discussion}

We have exhibited two tools---the Mahalanobis distance and fit to normal-mode $p_k$ values---that seem valuable and relatively robust for the purpose of testing and validating GP kernel choices.  As we have argued in this paper, such validation is of the essence if the probabilistic predictions of GP models are to be taken seriously, as is required in UQ applications of GP models.  It is surprising how infrequently one encounters concern for this point in the UQ literature. The work of Schonlau et al. \citep{schonlau1998global} contains the very useful suggestion of Leave-One-Out Cross Validation (LOO-CV) as a kernel validation method.  The methods presented here may be viewed as generalizations of LOO-CV, in that they make more general and higher-power use of the expected normal statistics of GP predictions.

We have made use of the Beta distribution as a simple family of distributions that can help discern deviations from uniformity.  This family is capable of representing deviations such as low-$p_k$ or high-$p_k$ biases, under-dispersion (a central ``hump''), and over-dispersion (``horns'' at low and high $p_k$) \citep[Chapter 8]{forbes2011statistical}.  These are commonly-encountered deviations from $\mathcal{U}(0,1)$, but they are not the only possibilities.  If exploration of histograms of $p_k$ should reveal, for example, two or more humps, or other such complex structures, it might be worth fitting a more complicated model to the data, such as a convex sum of Beta densities.  The proliferation of parameters (5 for a sum of 2 Beta densities) would probably mandate an MCMC analysis of the Bayesian posterior.

The examples of misspecified models given in the previous sections are somewhat simplistic, in that they represent the situation with respect to one-dimensional inputs and outputs.  In higher dimensions, there are opportunities for more general model misbehavior.  For example, correlations may arise between input dimensions that are poorly-modeled by popular kernel choices such as factored or isotropic kernels.  Additionally, vector-valued kernels for higher-dimensional outputs can be specified in a manner that cancels the correlations between different components, a phenomenon know as ``autokrigeability'' \citep{williams2007multi,alvarez2012kernels}.

These are the types of pathologies that afflicted early versions of the TAD algorithm, as briefly described in $\S$\ref{sec:TAD} above.  The problem was resolved by a brute-force approach in that case, with new similarly-structured additive kernel components compounded into the kernel model each time model inadequacy was detected by the Mahalanobis test.  There are certainly better, more efficient approaches than the brute-force strategy, but that strategy was adequate to the purpose of making the algorithm converge: the model merely needed to be good enough, and a far-from-perfect model served this purpose. Depending on the delicacy of one's needs, it might be necessary to attempt a more sophisticated approach to improving kernel fidelity. Examples of such sophisticated strategies include Deep Kernel Learning \citep{wilson2016deep} and Compositional Kernel Learning \citep{sun2018differentiable}, which are in turn based on kernel composition methods discussed in detail in \cite[][Chapter 4]{williams2006gaussian}. Nonetheless, the the Mahalanobis distance and fit to normal-mode $p_k$ values should provide a robust framework for testing and validation of such kernels.

One final important point is that in essentially all realistic use cases, kernel models will always be misspecified to some extent, and it is unrealistic to expect perfect fidelity from such models. The important issue in kernel validation is ensuring that normal predictions from GP models should be adequate to the specific purpose at hand.  The coverage guarantees of validated GP credible regions are limited to the type and input-space distribution of the held-out data used for validation: for example, spatially-sparse validation data can speak to longer length-scale functional data, but might not adequately validate the model for use with shorter length-scale predictions.  It is essential to have clarity on the model's use, when thinking about what it means for a GP kernel to be validated.

\section*{Acknowledgments}
This work was supported by the Office of Advanced Scientific Computing Research,
Office of Science, U.S. Department of Energy, under Contract DE-AC02-06CH11357.

\bibliographystyle{chicago}
\bibliography{refs}
\end{document}